\documentclass[twocolumn, pra]{revtex4-1}
\usepackage{amssymb}
\usepackage{amsmath}
\usepackage{epsfig}
\usepackage{color}
\usepackage{graphics, graphicx}
\usepackage{bbold}
\usepackage{psfrag}
\usepackage{mathcomp}
\usepackage{subfigure}
\usepackage{verbatim}
\usepackage{color}
\usepackage[colorlinks, citecolor=blue]{hyperref}

\begin{document}
\title{Universal One-dimensional Atomic Gases Near Odd-wave Resonance}
\author{Xiaoling Cui}
%\email{xlcui@iphy.ac.cn}
\affiliation{Beijing National Laboratory for Condensed Matter Physics, Institute of Physics, Chinese Academy of Sciences, Beijing 100190, China}
\date{\today}
\begin{abstract}
We show the renormalization of contact interaction for odd-wave scattering in one-dimension(1D). Based on the renormalized interaction, we exactly solve the two-body problem in a harmonic trap, and further explore the universal properties of spin-polarized fermions near odd-wave resonance using the operator product expansion method. It is found that the high-momentum distribution behaves as $C/k^2$, with $C$ the odd-wave contact. Various universal relations are derived. Our work suggests a new universal system emergent in 1D with large odd-wave scattering length.
\end{abstract}

\maketitle

\section{Introduction} 
%Odd-wave (or p-wave) interaction in one dimension (1D) is well known to host Majorana fermions\cite{Kitaev}, a typical topological quasi-particle that has attracted much attention in solid state physics in recent years\cite{Franks}. Ultra-cold gases emerges as ideal platform to .. as it can conveniently manipulate ... By using the confinement-induced-resonance\cite{CIR_p_expe,CIR_p_1,CIR_p_2,CIR_p_3}, ultra-cold gases provide a very convenient way to manipulate the strong odd-wave interaction in a variety of quasi-1D atomic systems, including spin-polarized fermions and various fermions/bosons mixtures\cite{...}, taking advantage of the p-wave Feshbach resonances in unconfined 3D gases\cite{}. Theoretical tools to study the 
A remarkable feature of cold atomic gases in the strong coupling regime is the universal property they exhibit. %One prominent example is the spin-1/2 Fermi gas across a broad Feshbach resonance, where the s-wave scattering length $a_s$ is much larger than the interaction range and serves as the unique interaction parameter of the system.
% uniquely determines the interaction effect in the system.
%the system is universally determined by the single interaction parameter $a_s$.
In this regime, a set of universal relations can be established to describe various microscopic and thermodynamic properties connected by a key quality called the {\it contact}, as first pointed out in a spin-1/2 Fermi gas near the s-wave Feshbach resonance\cite{Tan, Braaten, Zhang}. These relations provide a powerful understanding for the strongly interacting system and have been successfully verified in experiments\cite{Jin_contact1,Jin_contact2,contact3}.  Later the universal relations were also studied in other atomic systems such as bosons\cite{Braaten1}, in low-dimensions\cite{Werner, VZM, Zwerger,Valiente} and with higher partial-wave scatterings\cite{Ueda, Yu, Zhou}. Very recently the universal properties of a spin-polarized Fermi gas have been successfully explored near the p-wave Feshbach resonance\cite{Toronto}.

In this work, we point out a new system that exhibits universal properties, i.e., the one-dimensional (1D) atomic gases near odd-wave resonance. Such system can be realized~\cite{CIR_p_expe,CIR_p_1,CIR_p_2,CIR_p_3} by applying tight transverse confinements to 3D gases near p-wave Feshbach resonances, such as in identical fermions of $^{40}$K or $^{6}$Li\cite{K40, K40_2, Toronto, Li6_1,Li6_2} and in various atomic mixtures\cite{K40_2, FR_review}.
%$^{40}$K or $^{6}$Li Fermi gases near p-wave Feshbach resonances\cite{K40, Toronto, Li6_1,Li6_2}, thereby realizing the odd-wave resonance in quasi-1D geometry\cite{CIR_p_1,CIR_p_2,CIR_p_3}.  %%%Here a potential advantage of 1D system is that it can be highly stable near the resonance, on contrary to the 3D system with severe atom loss near a p-wave resonance. This is facilitated by the absence of centrifugal barrier for the relative motion of two particles in 1D, and thus the shallow bound state is much more extended and less possible to decay to the deep molecules and causes losses. Such advantage enables the exploration of strong interaction effect near a p(odd)-wave resonance in a practically stable atomic system.
To study the strong odd-wave interaction effect in these systems, it is fundamentally important to construct a model potential for the pairwise short-range interaction. In literature, several types of contact potential have been proposed\cite{Sen1,Sen2,Grosse,Girardeau1,Girardeau2}, which are all equivalent to the following form\cite{note_form}:
%Previous studies on such system have proposed a contact potential to simulate the short-range interactions\cite{..}:
%In literature there have been a number of studies on the spin-polarized fermions in 1D. ln particular, a contact potential is proposed to simulate the short-range interactions\cite{..}:
\begin{equation}
U(x)=U \overleftarrow{\partial}_x \delta(x) \overrightarrow{\partial}_x,  \label{contactU}
\end{equation}
where $x$ is the relative coordinate of two atoms with mass $m$, and $U=2l_o/m$ denotes the coupling strength proportional to the odd-wave scattering length $l_o$. In this paper we set $\hbar=1$. Despite the simple form of (\ref{contactU}), we will point out that such a potential with coupling $U\propto l_o$ is {\it unrenormalized}. %an inevitable defect of such model potential with coefficient $U \propto l_o$, i.e., it is not renormalizable.
It will produce ultraviolet divergence in the basic two-body scattering process in momentum space(see Fig.1), as already indicated in the second-order perturbation calculations\cite{Sen1,Sen2,Blume,Gora}. This will lead to unphysical result associated with short-range physics. Though special techniques can be employed in real space to avoid such problem for two-particle system\cite{Sen1, Blume2},
%such problem can be fixed in a two-particle system by the point-splitting prescription\cite{Sen1, Blume2},
%i.e., deliberately avoiding the $x=0$ point in (1),
it is not clear how these techniques work practically for many particles. Thus for a general many-body setting, Eq.(\ref{contactU}) with $U \propto l_o$ is likely to approximate the weak coupling limit giving the Hartree-Fork interaction energy\cite{Cui, Chen, Pu}, but not the strong coupling regime where the high-momenta scatterings are essential. In fact, previous rigorous studies on spin-polarized fermions, including the Bethe-ansatz solutions\cite{BA1,BA2} and the theorem of Bose-Fermi duality and its applications\cite{duality, Bender}, have utilized the boundary condition instead:
\begin{equation}
\lim_{x\rightarrow 0^{\pm}}\frac{\psi'}{\psi}=\mp \frac{1}{l_o},  \label{BP}
\end{equation}
here $\psi$ is the wave function and  $\psi'\equiv \partial \psi /\partial x$. Nevertheless, there have been rare discussions on the renormalization of 1D odd-wave interaction\cite{Nikolaj}, which is crucially important for the study of strong coupling regime with large odd-wave scattering length.

%and in particular, the potential (\ref{contactU}) has never been used to deal with the strong coupling regime with large odd-wave scattering length.

\begin{figure}[t]
\includegraphics[width=8.5cm]{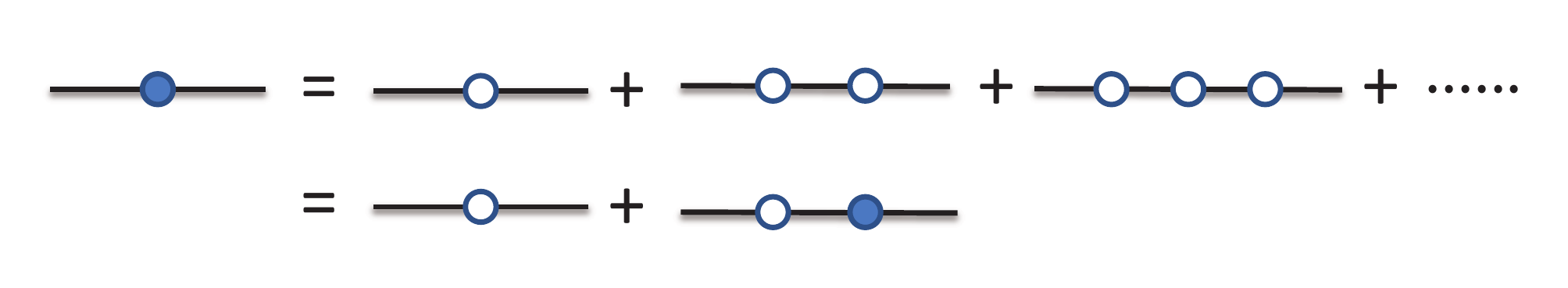}
\caption{(Color online). Diagrams for interaction renormalization. The solid (hollow) circle represents the effective (bare) interaction $T$ $(U)$. The lines denotes the incident and outgoing scattering states for the relative motion of two particles, given that it is separable from the center-of-mass motion. In continuum the state can be characterized by wave vector  $k$. % for continuum or the discrete level index $n$ for trapped system.
} \label{fig:schematic}
\end{figure}

With above motivations, in this work we first renormalize the contact potential in Eq.(\ref{contactU})  %for odd-wave scattering in 1D. This is done
by recognizing that $U$ therein is the bare coupling rather than the effective one($=2l_o/m$). Importantly, $U$ is related to $l_o$ by the renormalization equation:
\begin{equation}
\frac{1}{U}=\frac{m}{2l_o}-\frac{1}{L}\sum_k \frac{k^2}{2\epsilon_k},  \label{RG}
\end{equation}
with $\epsilon_k=k^2/(2m)$ and $L$ the length of the system\cite{footnote_L}. As an application, we exactly solve the two-body problem in a harmonic trap across the odd-wave resonance and demonstrate the Bose-Fermi duality\cite{duality} in the two-body version. We also show that (\ref{RG}) correctly produces the boundary condition (\ref{BP}) in both homogeneous and trapped cases. 
% the . At zero effective-range the two-body spectrum  produces  the even-wave interacting case with scattering length $l_e$ of the same value of $l_o$, consistent with the theorem of Bose-Fermi duality\cite{duality}.
% present the energy spectrum as well as the wave functions.
Compared with the boundary condition method, the renormalized potential has unique advantage in %offers a unique and efficient tool for 
addressing the momentum-space correlations in strongly interacting systems and exploring universal properties therein. In combination with the quantum field approach of operator-product-expansion, we further derive various universal relations for spin-polarized 1D Fermi gases near odd-wave resonance. In particular, the odd-wave contact, $C$, is identified in the high-momentum distribution as $\rho(k)\rightarrow C/k^2$ to the leading order. Our results can be straightforwardly generalized to other 1D systems with spin degree of freedom or with finite effective range, where the Bethe-ansatz method or Bose-Fermi duality could fail to work.
  
The rest of the paper is organized as follows. In section II, we present the interaction renormalization for 1D odd-wave scattering. In section III, we apply the renormalized contact potential to solve the two-body problem in a harmonic trap, with zero and finite effective range. In section IV we derive various universal relations of spin-polarized fermions by which the odd-wave contact $C$ is defined. %using the operator-product-expansion method.
The experimental relevance of our results is discussed in section V, and finally section VI is contributed to the summary and the outlook.

\section{Interaction Renormalization}

We start by showing the necessity of interaction renormalization for odd-wave scattering in 1D. Consider two atoms scattering with incident relative momentum $k$ and outgoing $k'$, the bare coupling $U_{k'k}\equiv \langle k'|U|k\rangle=(U/L) k'k$, while the scattering matrix $T$ is obtained by summing over all diagrams shown in Fig.1, giving (see Appendix A) 
\begin{equation}
%T(k,k')=U(k,k')+U(
\frac{k'k}{T_{k'k}(E)}=\frac{L}{U}-\sum_q \frac{q^2}{E-q^2/m+i0^+},  \label{RG2}
\end{equation}
here $E=k^2/m$ is the incident energy. Apparently, the second term in Eq.(\ref{RG2}), caused by all virtual scattering processes in Fig.1, has ultraviolet divergence proportional to the cutoff momentum $\Lambda$. As $T$ is a physical quantity describing effective interaction in the low-energy space irrelevant to short-range (or high-momentum) physics, the only way to satisfy Eq.(\ref{RG2}) is to require $1/U$ have the same divergence ($\sim\Lambda$), %Otherwise, if $U$ is finite ($\propto l_o$), $T$ will be unphysically related to the high-momentum cutoff or short-range details.
and that is how $U$ satisfies the renormalization equation (\ref{RG}). Note that this is in sharp contrast to the even-wave case, where such ultraviolet divergence is absent and the coupling constant needs not to be renormalized.

Given Eqs.(\ref{RG},\ref{RG2}), the $T$-matrix can be expressed by a few physical parameters:
\begin{equation}
\frac{k'k}{T_{k'k}(E)}=\frac{mL}{2}  \left(\frac{1}{l_o}+i k \right). \label{T-lp}
\end{equation}
%and together with Eq.\ref{RG2}, we will then get the renormalization equation (\ref{RG}), which connect the bare coupling $U_p$ with the odd-wave scattering length $l_p$.
By examining the pole of $T$-matrix element, we see a two-body bound state emerge when $l_o$ crosses resonance from $-\infty$ to $+\infty$ with the binding energy 
\begin{equation}
\epsilon_{b}=-\frac{1}{ml_o^2}. \label{Eb}
\end{equation}

To gain more physical understanding of $l_o$ in
%We show below that the quantity $l_o$ as defined in ... is consistent with the boundary condition...
Eqs.(\ref{RG},\ref{T-lp}), we study the two-body scattering wave function from the Lippman-Schwinger equation:
\begin{equation}
\psi(k')=2\pi \delta_{k,k'} + (E-k'^2/m+i0^+)^{-1} T_{k'k}(E). \label{psi_k}
\end{equation}
After Fourier transformation, we obtain the real-space wave function as (see appendix A)
\begin{equation}
\psi(x) \propto \left\{\begin{array}{l} \sin(kx+\delta),\ \ \ \,\,\,\,  (x>0); \\
\sin(kx-\delta),\ \ \ \,\,\,\,  (x<0), \end{array}\right. \label{wf}
\end{equation}
where the phase shift $\delta$ satisfies
\begin{equation}
\tan\delta=-kl_o. \label{Delta}
\end{equation} 
At scattering resonance $l_o\rightarrow \infty$, the phase shift saturates at $\delta=\pi/2$. Clearly the wave function satisfies the boundary condition (\ref{BP}). To this end, we have shown that the two definitions of scattering length $l_o$ in Eq.(\ref{BP}) and Eq.({\ref{RG}) are consistent.
%As tuning $1/l_p$ across the resonance from $0^-$ to $0^+$, a bound state will emerge with binding energy $E_{2b}=-2/(ml_p^2)$.

\begin{figure}[t]
\includegraphics[width=9cm]{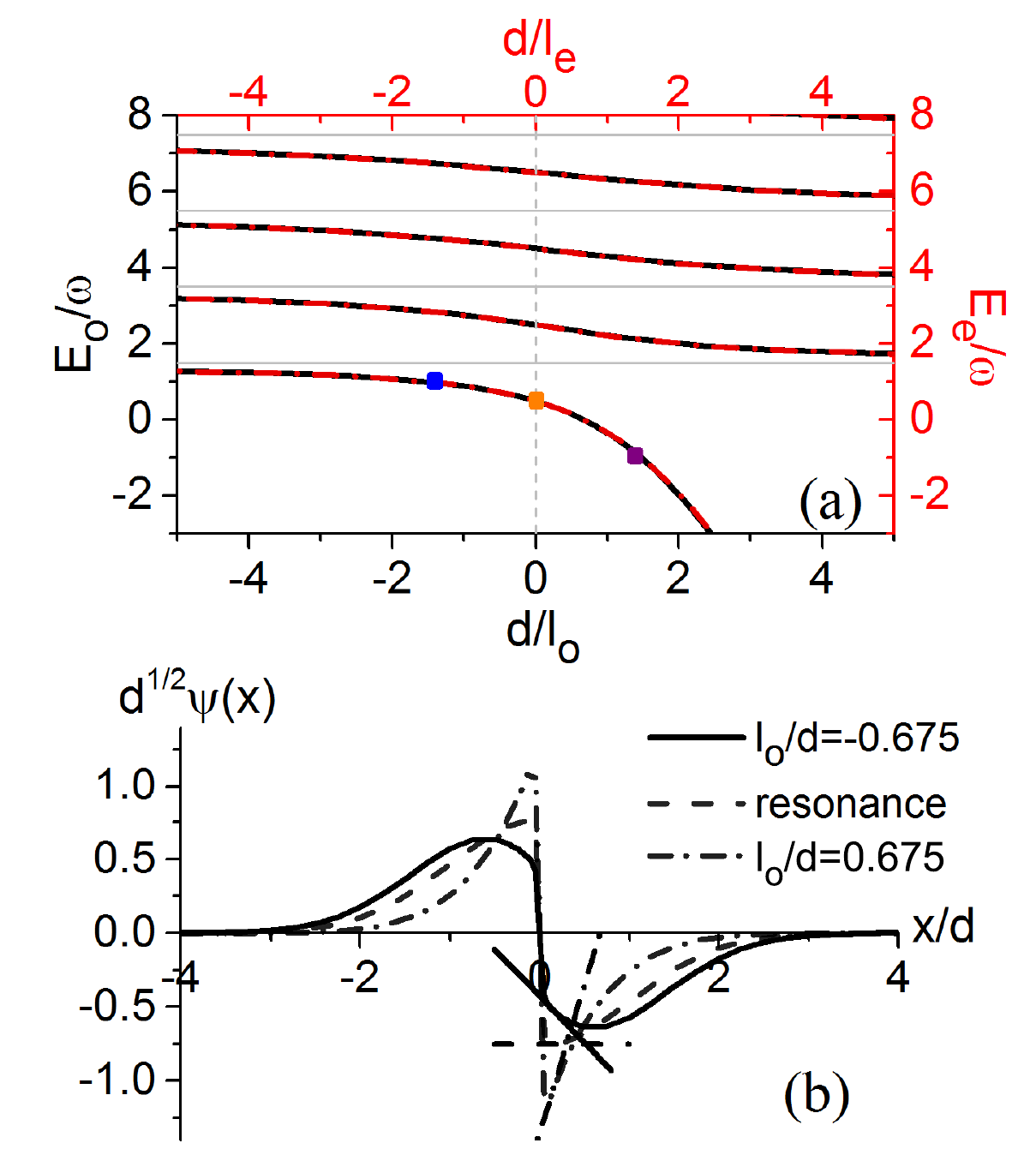}
\caption{(Color online). (a) Energy spectrum for the relative motion of two atoms in a harmonic trap with odd-wave ($E_o$-$l_o$, black solid) and even-wave ($E_e$-$l_e$, red dashed) interactions. $\omega$ and $d(=\sqrt{2/(m\omega)})$ are respectively the trap frequency and confinement length. Horizontal gray lines show the non-interacting odd-wave or the hard-core even-wave energy levels $(2l+3/2)\omega$ (here $l=0,1...$). While the vertical gray line at $l_o=l_e=\infty$ crosses the spectrum with energies $(2l+1/2)\omega$.
%The $E_o-l_o$ spectrum (black solid) is identical to $E_e-l_e$ spectrum (red dashed) for s-wave interacting case.
%(b) The odd- (black solid) and even-wave(red dashed) spectra for the lowest two branches with finite effective ranges $\xi_o=0.1md, \ \xi_e=-0.1md^3$. For comparison, gray lines show the spectra with zero range (same as in (a)). 
(b) normalized wave function $\psi(x)$ for odd-wave scattering at three scattering lengths $l_o/d=-0.675$(solid), $\infty$(dashed), and $0.675$(dash-dotted), as marked by squares from left to right in (a).  The according linear fits (with same line style) at $x\rightarrow 0^+$ confirm the boundary condition (Eqs.(\ref{BP},\ref{BP2})), i.e., $\psi(x\rightarrow 0^+)\propto (x-l_o)$.
} \label{fig2}
\end{figure}

\section{Two-body problem in a harmonic trap} 

We now apply the renormalized contact potential to solve the two-body problem in a harmonic trap, which was studied before using a different method\cite{Blume2}. Due to the separation of center-of-mass from relative motions, we only consider the Hamiltonian for the relative motion: $H=H_0+U$, with $H_0=-\nabla_x^2/m + (m/4)\omega^2 x^2$.
%\begin{equation}
%H_0=-\frac{1}{m}\frac{\partial^2}{\partial x^2}+\frac{1}{4}m\omega^2 x^2.
%\end{equation}
The eigen-wavefunction can be written as $\psi(x)=\sum_n c_n \phi_n(x)$, where $\phi_n(x)$ is the eigenstate of $H_0$ with energy $E_n=(n+1/2)\omega$ $(n=0,1...)$. By imposing the Schrodinger equation $H\psi=E\psi$, and utilizing the renormalization equation (\ref{RG}), we arrive at a closed equation for solving $E$ (see Appendix B):
\begin{equation}
\frac{m}{2l_o}=\sum_n \frac{|\phi_n'(0)|^2}{E-E_n} + \frac{1}{L}\sum_k \frac{k^2}{2\epsilon_k}.  \label{E_trap}
\end{equation}
One can check that the two terms in the right side of above equation produce  the same amplitude of ultraviolet divergence, thus can be exactly cancelled to ensure a physical solution of $E$. In practice, one needs to impose an energy cutoff for both terms, while the energy solution is insensitive to specific (large) cutoffs set.

In Fig.2(a), we plot the energy solution $E\equiv E_o$ as a function of $l_o$. It is found that $E_o$ approaches to $(2l+3/2)\omega$ and $(2l+1/2)\omega$ (here $l=0,1..$) respectively in the non-interacting ($l_o\rightarrow0$) and resonance ($l_o\rightarrow\infty$) limits. Such a spectrum is identical to the even-wave interacting case\cite{Busch}, with $l_o$ replaced by the even-wave scattering length $l_e$. %In fact, these two cases produces the same simplified equation for the spectrum. 
Physically, this is the two-body version of ``Bose-Fermi duality"\cite{duality}, stating that identical fermions can be mapped to bosons with reversed role of coupling strengths. However, such duality will break down when considering a finite effective range $\xi_{o/e}$, where $l_{o/e}$ becomes energy-dependent:
\begin{equation}
1/l_o \rightarrow 1/l_o+\xi_o E,\ \ \ \ \ l_e \rightarrow l_e+\xi_e E;
\end{equation}
Here the finite $\xi_o,\ \xi_e$ naturally arise when reducing the 3D interaction to quasi-1D geometry\cite{CIR_p_1,CIR_p_2, CIR_p_3,Olshanii, Cui2}. In Fig.3, we present the spectra at finite $\xi_{o}>0$ for odd-wave\cite{CIR_p_2} and 
$\xi_e<0$ for even-wave\cite{Cui2}. In comparison with the zero-range spectra, we can see that the finite effective ranges modify the spectra most prominently in the strong coupling regime ($l_o\rightarrow\infty,\ l_e\rightarrow 0$). 
while take little effects in the weak coupling limit ($l_o\rightarrow 0,\ l_e\rightarrow \infty$).

\begin{figure}[h]
\includegraphics[width=8.5cm]{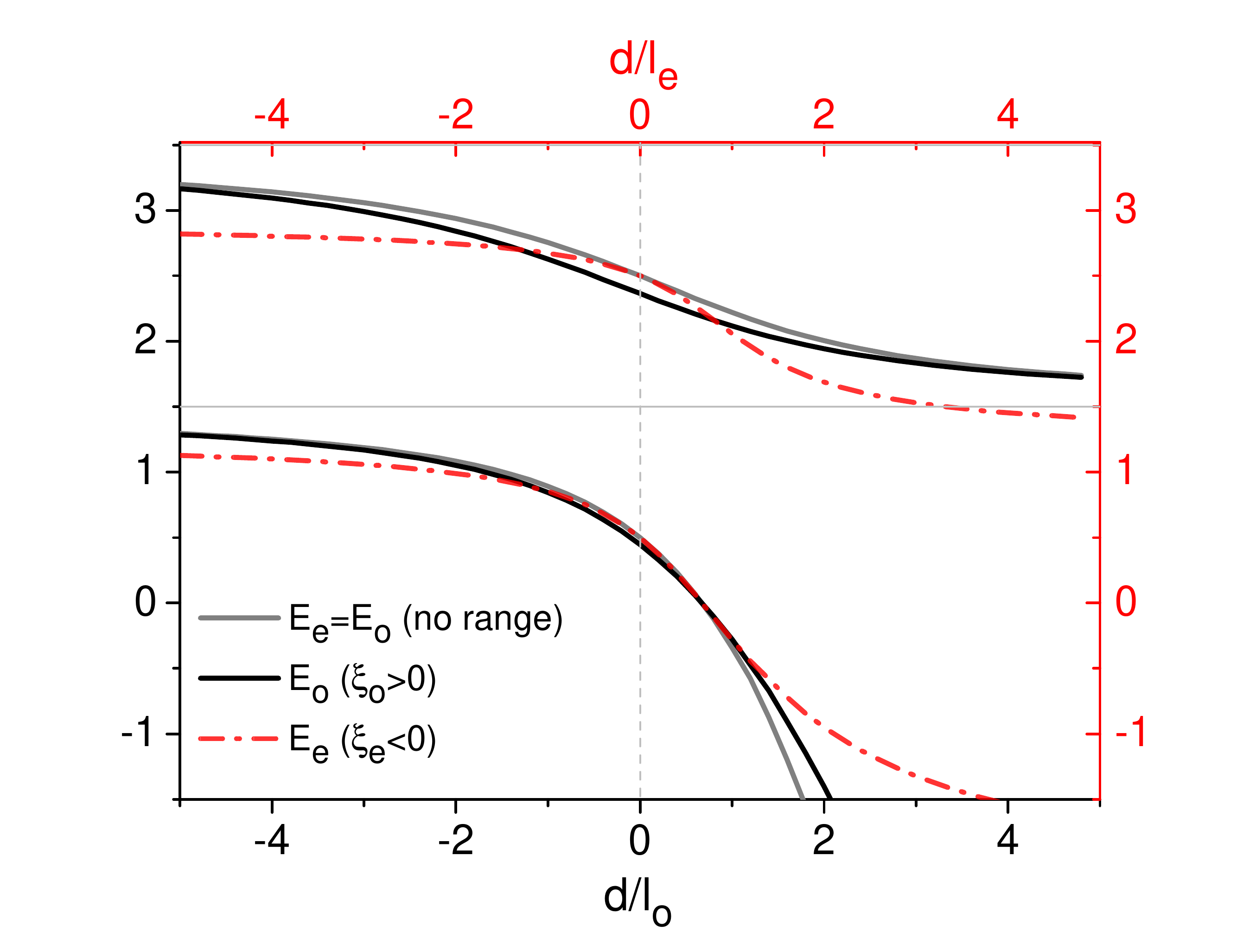}
\caption{(Color online). The odd- (black solid) and even-wave(red dashed) two-body spectra for the lowest two branches with finite effective ranges $\xi_o=0.1md, \ \xi_e=-0.1md^3$ ($d$ is the confinement length). For comparison, gray lines show the spectra with zero range (same as in Fig.2(a)). 
} \label{fig3}
\end{figure}

We now examine the wave function for odd-wave scattering (up to a normalized factor, see Appendix B):
\begin{equation}
\psi(x)=\sum_n \frac{\phi'^*_n(0)\phi_n(x)}{E-E_n}.  \label{wf_trap}
\end{equation}
The fact that only odd $n$ can contribute to the summation implies the odd-parity of $\psi$; meanwhile, the divergence of $\psi'$ at $x=0$ (as indicated by Eq.\ref{E_trap}) implies the discontinuity of $\psi$ when $x$ approaches zero from different sides. In Fig.2(b), we plot $\psi(x)$ at several typical values of $l_o$, which are found to well match the boundary condition (\ref{BP}). In fact, by utilizing Eq.\ref{E_trap}, we can prove the following asymptotic behavior of $\psi$ (see Appendix B):
\begin{equation}
\lim_{x\rightarrow 0} \frac{\psi(x)}{x} \rightarrow A (\frac{1}{l_p}-\frac{1}{|x|}), \label{BP2}
\end{equation}
which is an alternative expression of (\ref{BP}) while unifies the cases of $x\rightarrow 0^+$ and $x\rightarrow 0^-$ into a single compact form.

%The right-hand-side of Eq.(\ref{BP2}) shares remarkable similarity with the asymptotic behavior of 3D wave function at short range, where $l_o$ and $|x|$ are respectively replaced by the 3D s-wave scattering length $a_s$ and the inter-particle distance $r$. Such similarity indicate strongly the universality of 1D system at odd-wave resonance $l_p=\infty$.

\section{Universal relations} 

In the following, we derive the universal relations using the operator-product-expansion(OPE) for quantum fields\cite{Wilson, Kadanoff}, which has been successfully applied to strongly interacting atomic gases in recent years\cite{Braaten, Zwerger}. For brevity, we consider spin-polarized fermions with interaction $H_{\rm int}=\frac{U}{2} \int dR {\cal V} (R)$, where
\begin{eqnarray}
{\cal V}(R)&=& \int dx \Psi^{\dag}(R+x/2)\Psi^{\dag}(R-x/2) \overleftarrow{\partial}_x \delta(x) \overrightarrow{\partial}_x \nonumber\\
&&\ \ \ \ \ \ \ \Psi(R-x/2)\Psi(R+x/2),  \label{h_int}
\end{eqnarray}
here $\Psi^{\dag}, \Psi$ are the field operators of fermions. %We will neglect the effective range at the moment.

{\it (i) High-momentum distribution}. We first address the large-$k$ tail of the momentum distribution $\rho(k)$, which is given by:
\begin{equation}
\rho(k)=\int dR \int dx e^{-ikx} \langle \Psi^{\dag}(R-\frac{x}{2}) \Psi(R+\frac{x}{2})\rangle.  \label{rho_k}
\end{equation}
Since the large-$k$ behavior of $\rho(k)$ is essentially determined by the one-body density matrix
%(at right-hand-side of Eq.$\langle \psi^{\dag}(R-\frac{x}{2}) \psi(R+\frac{x}{2}) \rangle$
at short distance $x\rightarrow 0$, we can utilize the OPE for expansion: %expand the operator:
\begin{equation}
\Psi^{\dag}(R-\frac{x}{2}) \Psi(R+\frac{x}{2})=\sum_n C_n(x) {\cal O}_n(R). \label{ope}
\end{equation}
Here the local operator ${\cal O}_n(R)$ can be constructed by quantum fields and their derivatives; the short-distance coefficient $C_n(x)$ can have non-analytic dependence on $x$, leading to power law tail of $\rho(k)$ according to Eq.(\ref{rho_k}).

In the following, we will extract the leading non-analyticity in $C_n(x)$ and its according local operator ${\cal O}_n(R)$.
%show that the leading non-analytic term in $C_n(x)$ is given by $\sim |x|$, {\color{red} and the according local operator ${\cal O}_n(R)$ is just ${\cal V}(R)$ (Eq.\ref{h_int}).
As (\ref{ope}) is an operator equation, we can determine $C_n(x)$ by calculating the expectation value of each operator in the simplest quantum state. The state is chosen as $|\pm q\rangle$\cite{Braaten,Zwerger}, describing two colliding fermions with momenta $q$ and $-q$ and with total energy $E=q^2/m$. It is also convenient to define the amplitude of two-body scattering vertex as  (see Fig.4(a))
\begin{equation}
A(E)=-\frac{T_{q'q}(E)}{q'q}=-\frac{2}{mL} \left(\frac{1}{l_o}+iq\right)^{-1}. \label{A_E}
\end{equation}
%, here the factor of ``2" is due to the anti-symmetrization of fermionic wave functions\cite{supple}.

\begin{figure}[t]
\includegraphics[width=9cm]{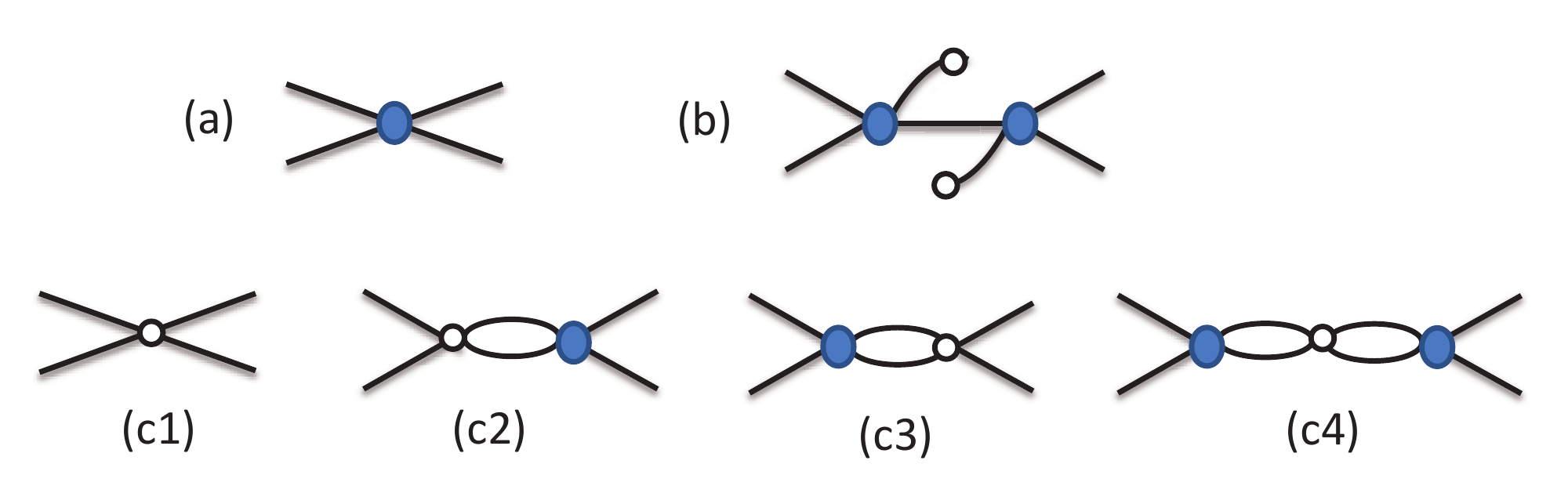}
\caption{(Color online). (a) Two-body scattering vertex $i A(E) qq'$. Here $q$ ($q'$) is the incident  (outgoing) momentum of the relative motion; $E=q^2/m$. (b) Diagram producing non-analyticity in $\langle \psi^{\dag}(R-\frac{x}{2}) \psi(R+\frac{x}{2}) \rangle$. The two open dots represent respectively the operators $ \psi^{\dag}(R-\frac{x}{2})$ and $\psi(R+\frac{x}{2})$.  (c1-c4) Four diagrams contributing to $\langle {\cal V}(R) \rangle$. Each open dot connecting with four lines represents the local operator ${\cal V}(R)$.%The open dots represent the operators.
} \label{fig4}
\end{figure}

The left side of OPE equation (\ref{ope}) can produce four types of diagrams, depending on whether the incoming and outgoing fermion lines are directly connected to the operators or through a scattering vertex in-between. Three of the diagrams produce analytic functions of $x$, which can be matched by matrix elements of one-body local operator $\psi^{\dag}\psi(R)$ and its derivatives. %The leading non-analyticity comes from 
The only non-analyticity comes from the diagram shown in Fig.4(b), which contains two scattering vertices and includes an integral of momentum flowing between operators and  vertices and between two vertices, similar to s-wave cases\cite{Braaten, Zwerger}. Explicitly, Fig.4(b) gives 
\begin{widetext} 
\begin{eqnarray}
\langle \psi^{\dag}(R-\frac{x}{2}) \psi(R+\frac{x}{2}) \rangle &=&q^2 (2iA(E))^2 i^3 \int \frac{dp dp_0}{(2\pi)^2} \frac{p^2 e^{ipx}}{(p_0-p^2/(2m)+i0^+)(E-p_0-p^2/(2m)+i0^+)^2} \nonumber \\
&=&m^2 A(E)^2 i|q|(1+i|q||x|)e^{i|q||x|}. \label{3b}
\end{eqnarray}
\end{widetext}
Its leading non-analytical term is $-2q^2m^2A(E)^2|x|$. More details regarding the Feynman diagrams and the derivation of Eq.(\ref{3b}) can be found in Appendix C.

As the one-body operators and their derivatives cannot produce such non-analyticity, one has to find more complicated local operators ${\cal O}(R)$ to match this term. Further by realizing that the non-analyticity comes from integration region with large momenta flowing, a natural choice of ${\cal O}(R)$ is the two-body interaction operator ${\cal V}(R)$, which corresponds to shrinking all internal lines in Fig.4(b) into a single point\cite{Braaten}.
%in the right side of OPE (\ref{ope}), such that its expectation value under  $|\pm q\rangle$ is proportional to $A(E)^2$. This turns out to be the vertex operator ${\cal V}(R)$ in (\ref{h_int}). {\color{red}It corresponds to shrinking all lines with large momentum flowing in Fig.3(b) into a single point\cite{Braaten}.} 
The relevant diagrams of ${\cal V}(R)$ are shown in Fig.4(c1-c4), which give:
\begin{eqnarray}
&(c1)&: \ \ \langle {\cal V}(R) \rangle=4q^2/L^2; \nonumber\\
&(c2)&: \ \ \langle {\cal V}(R) \rangle=4q^2 {\cal F}(E)/L^2; \nonumber\\
&(c3)&: \ \ \langle {\cal V}(R) \rangle=4q^2 {\cal F}(E)/L^2; \nonumber\\
&(c4)&: \ \ \langle {\cal V}(R) \rangle=4q^2 {\cal F}(E)^2/L^2; \nonumber
\end{eqnarray}
with the function ${\cal F}(E)$ given by
\begin{eqnarray}
&&\int \frac{dp dp_0}{(2\pi)^2} \frac{(i^3 A(E) L)\ p^2}{(p_0-p^2/(2m)+i0^+)(E-p_0-p^2/(2m)+i0^+)} \nonumber\\
&=&mA(E)L\big(\frac{\Lambda}{\pi}+\frac{iq}{2}\big),\label{fig3c}
\end{eqnarray}
where $\Lambda$ is the momentum cutoff. 
%The sum of all diagrams gives:
%\begin{equation}
%{\cal V}(R)=\frac{4q^2}{L^2} (1+{\cal F}(E))^2=4q^2 A(E)^2 \big(\frac{1}{A(E)L}+\frac{m\Lambda}{\pi}+\frac{imq}{2}\big)^2.\label{fig3c_2}
%\end{equation}
In combination with the expression of $A(E)$ in (\ref{A_E}) and the renormalization equation (\ref{RG}), finally ${\cal V}(R)$ can be reduced to
\begin{equation}
{\cal V}(R)=\frac{4q^2 A(E)^2}{U^2}.
\end{equation}
Thus we will have $(-|x|/2)m^2U^2\langle{\cal V}(R)\rangle$ in the right side of OPE equation (\ref{ope}) to match the leading non-analytical term in its left side. By denoting the contact as:
\begin{equation}
C=m^2 U^2 \int dR \langle {\cal V}(R) \rangle,
\end{equation} 
and in combination with Eq.(\ref{rho_k}), we arrive at the high-momentum distribution:
\begin{equation}
\rho(k)=\frac{C}{k^2}. \label{momentum}
\end{equation}
It is interesting to note that here $\rho(k)$ decays much more slowly than the even-wave case($\sim1/k^4$\cite{Zwerger}) in the high-$k$ regime. Physically this is because the pairwise wave-function at short-range is more singular in odd-wave(see Fig.2b) than in even-wave cases. 
%In the realistic quasi-1D atomic systems, Eq.(\ref{momentum}) can be verified in the region $k_F\ll k \ll 1/a_{\perp}$, where $k_F$ is the Fermi momentum and $a_{\perp}$ is the characteristic length of transverse confinements\cite{footnote_rhok}. %Next we derive a set of universal relations following the procedure in s-wave cases\cite{...}.

%{\color{red} Following the strategy of deriving universal relations for s-wave interacting case\cite{...}, we can  }

{\it (ii) Energy relation.} As the interaction energy $\langle H_{\rm int}\rangle$ is simply given by $C/(2m^2 U)$, %in combination with Eq.(\ref{RG}), 
we obtain the total energy:
\begin{equation}
E=\int \frac{dk}{2\pi} \epsilon_k \left(\rho(k)-\frac{C}{k^2}\right) +\frac{C}{4ml_o}+\langle V_T\rangle,
\end{equation}
with $\langle V_T\rangle$ the trapping energy. Here the presence of $-C/k^2$ in the bracket ensures the cancellation of ultraviolet divergence in the kinetic energy. Similar cancellation also occurs in 2D and 3D s-wave cases\cite{Tan, Braaten, Werner, VZM}.

{\it (iii) Adiabatic relation.} Using renormalization equation and Feynman-Hellman theorem\cite{Braaten, Tan, Zhang, Zwerger}, we obtain
\begin{equation}
\frac{\partial E}{\partial (-1/l_o)} = \frac{C}{4m}.  \label{adiabatic}
\end{equation}

{\it (iv) Virial theorem.} With a harmonic trapping potential $V_T=\sum_i m\omega^2 x_i^2/2$, one can derive the Virial theorem using dimensional analysis\cite{Tan, Braaten, Zhang, Zwerger}, which requires $(\omega \partial/\partial \omega -  l_o/2 \partial/\partial l_o)E=E$. %Using the  Feynman-Hellman theorem together with the adiabatic relation (\ref{adiabatic}), 
Finally we obtain
\begin{equation}
E = 2 \langle V_T\rangle-\frac{C}{8m l_o}.  \label{virial}
\end{equation}

{\it (v) Pressure relation.} In a homogeneous system, the pressure relation can be obtained by applying dimensional analysis\cite{Tan, Braaten, Zhang, Zwerger} to the free energy density ${\cal F}=F/L$, which requires $(T \partial/\partial T +n/2 \partial/\partial n - l_o/2 \partial/\partial l_o){\cal F}=3/2 {\cal F}$. Using Eq.(\ref{adiabatic}), the pressure density ${\cal P}$ can then be related to the energy density ${\cal E}$ and the contact density ${\cal C}$ as:
\begin{equation}
{\cal P}=2{\cal E}+\frac{{\cal C}}{4ml_o}.  \label{pressure}
\end{equation}

{\it (vi) Tail of rf spectroscopy.} Through a rf transition, the fermions can be transferred to an empty state by $H_{\rm rf}=\Omega\sum_k \psi_k^{e\ \dag} \psi_k$. The Fermi's Golden Rule gives the transition rate $\Gamma(\omega)=2\pi \sum_f |\langle f |H_{rf}|i\rangle|^2 \delta(\omega+E_i-E_f)$, with $i,f$ labeling the initial and final states and $E_i,\ E_f$ the corresponding energies. The high-frequency tail of the rf spectrum\cite{Punk, Schneider, Platter, Yu} can be obtained as:
\begin{equation}
\Gamma(\omega)=\frac{{\Omega}^{2}}{\sqrt{2m}} \frac{C}{\omega^{3/2}}. \label{rf}
\end{equation}

Facilitated by the Bose-Fermi duality\cite{duality}, the odd-wave contact of identical fermions ($C\equiv C_o$) can be related to the even-wave contact of identical bosons $C_e$ defined by $\partial E/\partial l_e=C_e/(4m)$\cite{Zwerger, Valiente}. At given $l_o=l_e\equiv l$, we have $C_o=C_e l^2$. One can thus infer $C_o$ near the odd-wave resonance of fermions from $C_e$ in the weak coupling limit of bosons, resulting in $C_o/L=4\rho^2$ for the homogeneous system ($\rho$ is the uniform density). The resulted $\rho(k)$ in Eq.(\ref{momentum}) is consistent with that found previously using a different method\cite{Bender, note}.

The universal relations derived here can be easily generalized to systems with spin degree of freedom, such as in f(fermion)-f, f-b(boson), and b-b atomic mixtures.
The finite range effect can also be included straightforwardly\cite{Braaten,Platter2}, and the related result will be presented elsewhere\cite{Cui3}.

%Table I:  a number of properties revealed in this system show close relation or similarity to those in the s-wave interacting 3D system, such as the renormalization equation, the boundary condition, the bound state and some of universal relations, as listed in Table I.

\section{Experimental relevance} 

Our results of the high-momentum distribution and various universal relations can be verified realistically in the p-wave interacting Fermi gas confined in quasi-1D geometry. In particular, the high-momentum distribution (\ref{momentum}) can be detected in the region $k_F\ll k \ll 1/a_{\perp}$, where $k_F$ is the Fermi momentum and $a_{\perp}$ is the characteristic length of transverse confinements. This condition can be satisfied either in a few-particle system in cigar-shaped traps with aspect ratio $\sim 10$\cite{science2011}, or in optical lattices confining 10-100 particles in each tube with aspect ratio$\sim$100-1000\cite{science2009}.

Another practically important factor for experimental detection is the atom loss. Hopefully near the odd-wave resonance the atom loss can be much suppressed compared to the 3D case near p-wave resonances\cite{Toronto, K40, K40_2, Li6_1,Li6_2}. Such suppression is indicated by the absence of centrifugal barrier for 1D collision and thus the spatially much extended bound state (see Fig.2b), which is less likely to decay into deep molecules and cause three-body losses. Moreover, the shifted resonance in quasi-1D\cite{CIR_p_1,CIR_p_2,CIR_p_3} also helps to avoid severe losses near the odd-wave resonance, whose location in magnetic field can be far from that of a 3D p-wave resonance.

\section{Summary and outlook}

In summary, this work presents the first analysis of interaction renormalization for the 1D odd-wave interacting atomic systems. Compared to previous approaches in dealing with  odd-wave interactions, such as the Bose-Fermi duality\cite{duality} (applied to spin-independent interaction and zero range) and the Bethe-ansatz methods\cite{BA1,BA2} (applied to spin-independent interaction), the renormalization approach has much broader applications to atomic systems, such as with spin-dependent interaction and with finite range. It also has unique advantage, compared to the boundary condition approach, in addressing the low-energy effective scattering and the momentum-space correlation. Based on the interaction renormalization, we derive various universal relations of the 1D spin-polarized Fermi gas, which can be verified in current cold atoms experiments.

The renormalization and the renormalized potential revealed in this work lay the foundation for exploring intriguing quantum phenomena in 1D induced by strong odd-wave interactions. In future, it is interesting to study how the Majorana fermion, proposed as a topological quasi-particle of electrons moving in quantum wires\cite{Kitaev}, can emerge in the 1D Fermi gas near odd-wave resonances. 
%, while Eqs.(\ref{contactU},\ref{RG}) obtained here provide the basic interaction model for such system. 

Finally, our work also reveals remarkable similarities between the 1D odd-wave interacting and 3D s-wave interacting spin-1/2 fermion systems in the two-body scattering   properties, such as the relation between phase shift and scattering length(Eq.\ref{Delta}), the emergence of a two-body bound state when the scattering length across resonance to $+\infty$ (Eq.\ref{Eb}), etc. Such two-body similarities may even result in many-body physics in common between these two systems. For instance, the spin-1/2 fermions in 1D with odd-wave interaction (between different species) may undergo a similar BCS-BEC crossover across the resonance as in the 3D s-wave case. This suggests an alternative approach to the rather challenging many-body problems in 3D fermions, i.e., by studying the 1D odd-wave counterpart. Hopefully our work will stimulate more studies along this route.

\bigskip

{\bf Acknowledgement.} I thank Selim Jochim, Paul Julliene, Zhenhua Yu, Doerte Blume, Tin-Lun Ho and Nikolaj Zinner for helpful discussions.
The work  is supported by the National Natural Science Foundation of China (NNSFC) under grant No.11374177, No. 11534014, and by the programs of Chinese Academy of Sciences.

\appendix

\section{Scattering matrix and phase shift}

The scattering matrix element $T_{k'k}(E)=\langle k' |T| k\rangle $ can be obtained by summing up all ladder diagrams shown in Fig.1:
\begin{widetext}
\begin{eqnarray}    
T_{k'k}(E)&=&U_{k'k} + \sum_{q} U_{k'q} \frac{1}{E-q^2/m+i0^+} U_{qk}+ \sum_{qq'} U_{k'q} \frac{1}{E-q^2/m+i0^+} U_{qq'}\frac{1}{E-q'^2/m+i0^+} U_{q'k} +... \nonumber\\
&=& \frac{k'k}{L}\left( U+U^2\frac{1}{L}\sum_q \frac{q^2}{E-q^2/m+i0^+}  + U^3 \big(\frac{1}{L}\sum_q \frac{q^2}{E-q^2/m+i0^+} \big)^2 +...\right); \label{fig1_1}
\end{eqnarray}
\end{widetext}
denoting $f=LT_{k'k}(E)/(k'k)$, then we get
\begin{eqnarray}    
f&=& U+\frac{U}{L}\sum_q \frac{q^2}{E-q^2/m+i0^+} f.   \label{fig1_2}
\end{eqnarray}
This gives Eq.(\ref{RG2}).

Given the momentum-space wave function (\ref{psi_k}), one can obtain the real-space form:
\begin{eqnarray}
\psi(x)&=&\frac{1}{\sqrt{L}} \sum_{k'} \psi(k') e^{ik'x}  \nonumber\\
&=& \frac{1}{\sqrt{L}}\left( i\sin kx +\int \frac{dk'}{2\pi} \frac{e^{ik'x}}{k^2-k'^2+i0^+} \frac{2k'k}{1/l_o+ik} \right) \nonumber\\
&=&\frac{1}{\sqrt{L}}\left( i\sin kx + (-i) sgn(x) e^{ikx} \frac{k}{1/l_o+ik} \right),
\end{eqnarray}
here we have anti-symmetrized the first term (incident wave function) to ensure the odd-parity of $\psi$; $sgn(x)=-1 \ ({\rm or}\  0) $ if $x<0 ({\rm or} >0)$. Denoting $\tan\delta=-kl_o$, the wave function can be reduced to  Eq.(\ref{wf}) (up to a factor), where $\delta$ is the phase shift. One can easily check the asymptotic behaviors of $\psi$ in the limit of $x\rightarrow 0$ as: $\psi(x\rightarrow 0^{\pm})\sim (x\mp l_o)$, consistent with the boundary conditions Eqs.(\ref{BP},\ref{BP2}) in the main text.

\section{Two-body problem in a harmonic trap}

\subsection{Derivations of Eqs.(\ref{E_trap},\ref{wf_trap})}

Two methods to derive Eqs.(\ref{E_trap},\ref{wf_trap}) are in order:

Method (1):  

Expand the wave function as $\psi(x)=\sum_n c_n \phi_n(x)$ and use the Schrodinger equation $H\psi=E\psi$, we can get 
\begin{equation}
(E-E_n) c_n=U\sum_m c_m \psi'^*_n(0)\psi'_m(0).
\end{equation}
Denote 
\begin{equation}
U\sum_m c_m \psi'_m(0)\equiv A, \label{eq1}
\end{equation}
we have 
\begin{equation}
c_n=\frac{A\psi'^*_n(0)}{E-E_n}. \label{eq2}
\end{equation}
Eqs.(\ref{eq1},\ref{eq2}) will then produce 
%we can easily obtain 
the self-consistent equation for $E$:
\begin{equation}
\frac{1}{U}=\frac{|\psi'_n(0)|^2}{E-E_n}.  \label{pole}
\end{equation}
Combining with the renormalization equation (\ref{RG}), we will arrive at 
Eq.(\ref{E_trap}). Given Eq.\ref{eq2}, we can also get the wave function expressed by Eq.(\ref{wf_trap}).

Method (2):

The bound state solution is determined by the pole of $T-$matrix element ($T_{nn'}$) between two arbitrary energy levels $n,n'$. As shown in Fig.1, $T_{nn'}$ can be obtained by summing up all relevant ladder diagrams. Following the similar derivation as in the continuum case, we can get $T_{n'n}=\phi'^*_{n'}(0)\phi_n(0) f$, and $f$ satisfies
\begin{eqnarray}    
f&=&U+U\sum_n \frac{|\psi'_n(0)|^2}{E-E_n} f,   
\end{eqnarray}
which gives
\begin{eqnarray}    
\frac{1}{f}&=&\frac{1}{U}-\sum_n \frac{|\psi'_n(0)|^2}{E-E_n}.   
\end{eqnarray}
It is then obvious that the pole of $T_{n'n}$ (or $f$) is determined by Eq.(\ref{pole}) as obtained previously. Consequently we can further arrive at Eq.(\ref{E_trap}).

According to the Lippman-Schwinger equation, the wave function is given by:
\begin{equation}
|\psi\rangle = \frac{1}{E-H_0} U |\psi\rangle. \label{LS}
\end{equation}
Given the definition of $U$-operator (Eq.\ref{contactU}), we can assume
\begin{equation}
\langle x|U |\psi\rangle =f \overleftarrow{\partial}_x \delta(x). \label{LS2}
\end{equation}
Then the real-space wave function can be obtained as
\begin{equation}
\psi(x)\equiv \langle x|\psi\rangle = f \sum_n \frac{\phi_n(x)\psi'^*_n(0)}{E-E_n} ,\end{equation}
which is just Eq.(\ref{wf_trap}) up to a constant factor.

In order to solve Eq.(\ref{E_trap}), we set an energy cutoff $E_{c}=N_c \omega$ to the two summations in it and reduce it to:
\begin{equation}
\frac{d}{l_o}=\frac{1}{\pi} \lim_{N_c\rightarrow\infty} \left( 4\sqrt{N_c} -\sum_{l=0}^{N_c}\frac{(2l+1)!!}{(2l)!!)}\frac{1}{l-E/(2\omega)+3/4} \right). \label{E_nume}
\end{equation}
Here $d=\sqrt{2/(m\omega)}$ is the confinement length. One can check that the two terms in the right side of above equation have the same divergence ($\sim N_c^{1/2}$) as the cutoff $N_c\rightarrow\infty$, thereby can cancel with each other and result in a physical solution for $E$. In Fig.2a in the main text, we show the numerical result of $E$ by solving Eq.(\ref{E_nume}), and meanwhile confirm that the solution is insensitive to the choice of $N_c$ as long as $N_c$ is large enough (we have checked $N_c$ from 2000 to 8000).

\subsection{Boundary condition}

In this subsection, we prove the boundary condition as shown by Eq.\ref{BP2} in the main text. Given the wave function (\ref{wf_trap}), we have
\begin{equation}
\lim_{x\rightarrow 0} \frac{\psi(x)}{x}= \sum_n \frac{\phi'^*_n(0)\phi'_n(x\rightarrow 0)}{E-E_n}.  \label{psi_bc}
\end{equation}
We will show the right side of above equation diverges as $-1/|x|$ in the limit of $x\rightarrow 0$. 

We start by considering the odd-wave scattering in free space. By using Eqs.(\ref{LS},\ref{LS2}), we can write down the scattered wave function as: (up to a factor)\begin{equation}
\psi_f(x)=\frac{1}{2\pi}\int_{-\infty}^{+\infty} dq \frac{-iq e^{iqx}}{E-q^2/m+i0^+} = -\frac{m}{2} sgn(x) e^{ikx},   \label{wf_F}
\end{equation}
where $E$ is the incident energy and $k=\sqrt{mE}$ the incident wave vector. In the limit of $E,k\rightarrow 0$ and $x\rightarrow 0$, we have
\begin{equation}
\lim_{x\rightarrow 0} \frac{\psi_f(x)}{x} = -\frac{m}{2} \frac{1}{|x|} . \label{equiv1}
\end{equation}
On the other hand, the integral in Eq.\ref{wf_F} is equivalent to $\frac{1}{2\pi}\int_{-\infty}^{+\infty} dq \frac{q \sin qx}{E-q^2/m+i0^+}$, and in the same limits as above, we have \begin{equation}
\lim_{x\rightarrow 0} \frac{\psi_f(x)}{x} = \frac{1}{2\pi}\int_{-\infty}^{+\infty} dq \frac{q^2}{-q^2/m}=-\frac{1}{L}\sum_q \frac{q^2}{2\epsilon_q}. \label{equiv2}
\end{equation}
Thus Eq.\ref{equiv1} and Eq.\ref{equiv2} can be considered equivalent to each other. 

For the trapped case, since Eq.\ref{E_trap} guarantees that Eq.\ref{psi_bc} has the same divergence as Eq.\ref{equiv2} (which is equivalent to Eq.\ref{equiv1}), we can combine all these equations and get:
\begin{eqnarray}
\lim_{x\rightarrow 0} \frac{\psi(x)}{x}&=& \sum_n \frac{\phi'_n(0)\phi'_n(x\rightarrow 0)}{E-E_n} \nonumber\\
&=& \frac{m}{2l_o}-\frac{1}{L}\sum_q \frac{q^2}{2\epsilon_q} \nonumber\\
&=& \frac{m}{2}(\frac{1}{l_o}-\frac{1}{|x|}). 
\end{eqnarray}
To this end we prove Eq.(\ref{BP2}). Above proof can be applied to an arbitrary type of trapping potentials.

\section{Operator-Produce-Expansion method(OPE) for odd-wave interacting fermions}

In this appendix, we first illustrate Feynman rules for the OPE diagrams, and then present  details using OPE to derive the high-momentum distribution of identical fermions. 

\subsection{Feynman rules}

We first give the Feynman rules for evaluating the diagrams in OPE. 

(1) {\it Two-momenta and loop integral.} Each atom line is characterized by two-momenta ($p_0,p$), with $p_0$ the energy and $p$ the momentum. The two-momenta of incoming and outgoing lines are constrained by the conservation of total momentum and total energy. If there exists two-momenta  ($p_0,p$) independent on the two-momenta of incoming and outgoing lines, they should be integrated over using $L/(2\pi)^2\int dp dp_0 $.

(2) {\it Propagators.}  Each internal line with two-momenta ($p_0,p$) is assigned a propagator factor $i/(p_0-p^2/(2m)+i0^+)$.

(3) {\it Scattering vertices.} Each scattering vertex is assigned a factor $i A(E)qq'$ (see Fig.3(a) in the main text). Here $q$ ($q'$) is the incident  (outgoing) momentum of the relative motion of two particles, and $E=q^2/m$ is the incident energy. 

\begin{figure}[h]
\includegraphics[width=9cm]{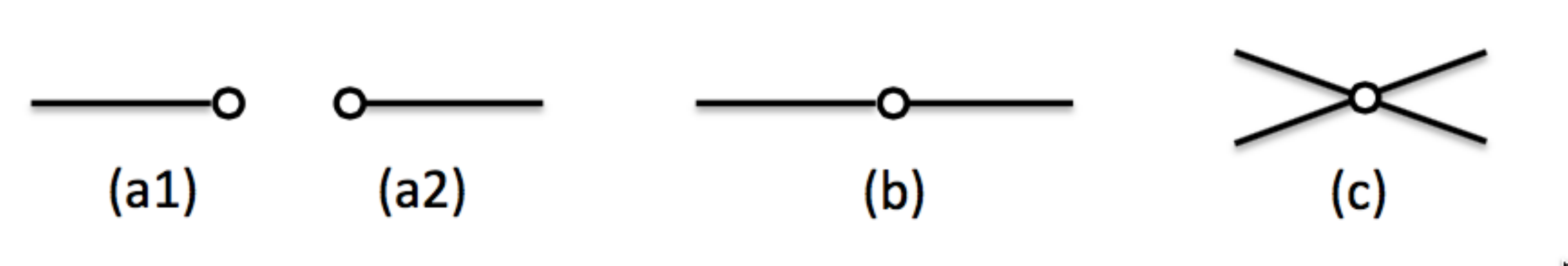}
\caption{(Color online). Diagrams of local operators: (a1) $\psi(R)$, (a2) $\psi^{\dag}(R)$; (b) one-body operators, such as $\psi^{\dag}(R)\psi(R)$ and its derivatives; (c) two-body operators, such as ${\cal V}(R)$.} \label{fig2s}
\end{figure}

(4) {\it Operator vertices.}

Two basic local operators $\psi(R)$ and $\psi^{\dag}(R)$ are diagrammatically shown in Fig.5(a1) and (a2). $\psi(R)$ (or $\psi^{\dag}(R)$) is denoted by an open dot with an atom line ending (or starting) at the dot. 

The one-body local operators, which annihilate one atom and create one atom at the same site, are diagrammatically shown in Fig.5(b). These operators include $\psi^{\dag}(R)\psi(R)$ and its derivatives $\psi^{\dag}(R)(\overleftarrow{\partial_R})^m\psi(R)$,  $\psi^{\dag}(R)(\overrightarrow{\partial_R})^m\psi(R)$ ($m$ is an integer). If the incoming and the outgoing momenta are $k$ and $k'$, these operators will produce:
\begin{eqnarray}
\psi^{\dag}(R)\psi(R): &&\ \ \ e^{i(k-k')R}/L ;\\
\psi^{\dag}(R)(\overleftarrow{\partial_R})^m\psi(R):&&  \ \ \ (-ik')^m e^{i(k-k')R}/L ;\\
\psi^{\dag}(R)(\overrightarrow{\partial_R})^m\psi(R):&& \ \ \ (ik)^m e^{i(k-k')R}/L.
\end{eqnarray}

The two-body local operators, which annihilate two atoms and create two atoms at the same site, are diagrammatically shown in Fig.5(c). A typical example is the interaction operator ${\cal V}(R)$. For incoming momenta $(k,-k)$ and outgoing momenta $(k',-k')$ of two identical fermions, such operator produces 
\begin{equation}
{\cal V}(R):\ \ \ \ 4kk'/L^2. \label{V_R}
\end{equation}
%This can be calculated by writing down the incident and outgoing wave-functions for two identical fermions. For instance, the incoming wave-function is
%\begin{eqnarray}
%\psi_{in}=\langle x_1,x_2|k,-k\rangle=\frac{1}{\sqrt{2}}\left( \phi_k(x_1)\phi_{-k}(x_2)-\phi_k(x_2)\phi_{-k}(x_1)\right)=\frac{\sqrt{2}}{L}i\sin kx;
%\end{eqnarray}  
%with $x=x_1-x_2$. Here we have used the single-particle wave-function $\phi_k(x)=e^{ikx}/\sqrt{L}$. Similarly the outgoing wave-function reads $\psi_{out}=\langle x_1,x_2|k',-k'\rangle=\frac{\sqrt{2}}{L}i\sin k'x$. One can easily get Eq.(\ref{V_R}) by evaluating the interaction matrix element between these two wave-functions.  

\subsection{OPE diagrams for identical fermions}

Evaluated under the two-particle scattering state $|\pm q\rangle$, the bi-local operator $\Psi^{\dag}(R-\frac{x}{2}) \Psi(R+\frac{x}{2})$ can involve four types of diagrams, as shown in Fig.6(a-d). In Fig.6(a), one incoming line (with momentum $q$) and one outgoing line (with the same $q$) are directly connected to the operators, which produces:
\begin{equation}
(a):\ \ \ \ e^{iqx}/L. \label{eq_a}
\end{equation}
In Fig.6(b) and (c), one (incoming or outgoing) line is connected to the operator through a scattering vertex. They both produce:
\begin{equation}
(b), (c):\ \ \ \ q^2 (2iA(E)) e^{iqx}/L. \label{eq_bc}
\end{equation}
In Fig.6(d) (equivalent to Fig.4b), both the incoming and outgoing lines are connected to the operators through scattering vertices. The associated matrix element involves an integral over the two-momenta of internal lines, giving Eq.(\ref{3b}) in the main text. Note that the factor of $2$ ahead of $iA(E)$ in the equations for Fig.6(b,c,d) is due to the fact that the scatterings are between two identical fermions and thus each scattering vertex $A(E)$ should be accompanied by a factor of 2. One can see this clearly by examining the expectation value of $U$(Eq.\ref{contactU}) under the state $|\pm q\rangle$, which is given by $2q^2 U/L$ instead of $q^2U/L$. 

\begin{widetext}

\begin{figure}[h]
\includegraphics[width=14cm]{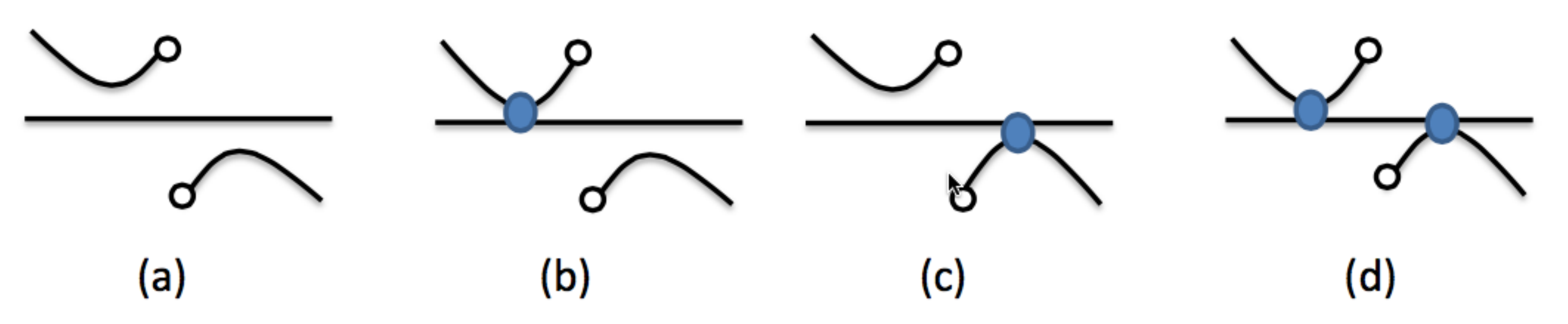}
\caption{(Color online). Diagrams for matrix elements of bi-local operator $\Psi^{\dag}(R-\frac{x}{2}) \Psi(R+\frac{x}{2})$ between the two-particle scattering states. Note that the diagram (d) is equivalent to Fig.4(b) in the main text.} \label{fig3s}
\end{figure}

\begin{figure}[h]
\includegraphics[width=14cm]{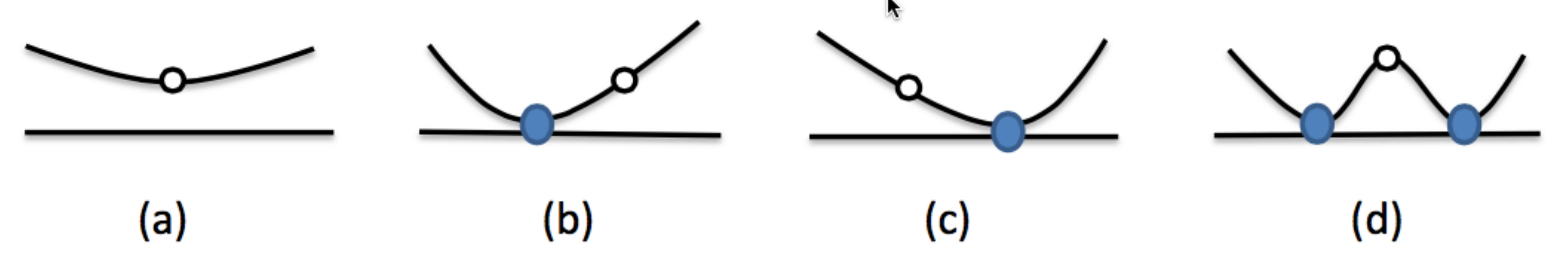}
\caption{(Color online). Diagrams for matrix elements of one-body local operators, such as $\Psi^{\dag} \Psi(R)$ and its derivatives, between the two-particle scattering states.} \label{fig4s}
\end{figure}

\end{widetext}

Next we will look for the local operators ${\cal O}_n(R)$ in the right side of OPE equation (\ref{ope})  to match the elements of $ \Psi^{\dag}(R-\frac{x}{2}) \Psi(R+\frac{x}{2})$  produced by the diagrams in Fig.6(a-d). First, we note that Eqs.(\ref{eq_a},\ref{eq_bc}) are all analytical in terms of variable $x$. Thus one can simply find all local operators by Taylor expanding these equations in terms of $x$. Take Fig.6(a) and the associated Eq.(\ref{eq_a}) for instance, $e^{iqx}=\sum_n (iq)^n x^n/n! \ (n=0,1,...)$, so the local operators are simply the one-body operators ${\cal O}_n(R)=\psi^{\dag}(\overrightarrow{\partial_R})^n\psi(R)$, and the coefficients are $C_n(x)=x^n/n!$. The according diagrams for those local operators are shown in Fig.7(a). Similarly, the diagrams in Fig.7(b) and (c) match the expansions in power of $x$ of the corresponding diagrams in Fig.6(b) and (c). The diagrams in Fig.7(d) match the even power of $x$ of the diagrams in Fig.6(d). However, Fig.6(d) also includes the odd power of $|x|$ (Eq.\ref{3b}), which cannot be matched by any diagram in Fig.7(a-d). These non-analytical terms  must be matched by more complicated local operators beyond the one-body ones.

Since the analytical functions of $x$ cannot produce any high-momentum tail in the Fourier expansion, we will only focus on the terms of odd power of $|x|$, which is non-analytic near $x=0$, produced by Eq.\ref{3b}. The leading non-analytic term in Eq.\ref{3b} is $-2q^2 m^2 A(E)^2 |x|$, which is linear in $|x|$ and gives rise to a high-momentum tail as $1/k^2$.  In order to match this term, one has to find a local operator ${\cal O}(R)$ such that its expectation value under  $|\pm q\rangle$ is proportional to $A(E)^2$. By realizing that the non-analyticity can only come from the integration region with large momentum flowing, and the fact that one should search for the operators beyond one-body ones, a natural choice of ${\cal O}(R)$ is the two-body interaction operator ${\cal V}(R)$. This corresponds to shrinking all internal lines in Fig.6(d) (or equivalently Fig.4(b)) into a single point. There are four relevant diagrams produced by ${\cal V}(R)$, as shown by Fig.4(c1-c4) in the main text.


\begin{thebibliography}{99}
\bibitem{footnote_L} Throughout the paper, for the continuum state we use the periodic boundary condition for atoms moving in a 1D tube of length $L$.
\bibitem{note} Note that the normalization of $\rho(k)$, $\mathcal{R}\equiv\int dk/(2\pi) \rho(k)$, is different between Ref.\cite{Bender} and our work: $\mathcal{R}$ is $N/(2\pi L)$ in \cite{Bender} while is $N$ in ours ($N$ is total particle number).
\end{thebibliography}
\end{document}